\documentclass[11pt,a4paper]{article}
\usepackage{geometry}
\usepackage[round]{natbib}
\usepackage{authblk}
\usepackage{graphicx}
\usepackage{tabularx}
\usepackage{multirow}
\usepackage{multicol}
\usepackage{amsthm}
\usepackage{algorithm2e}
\usepackage{ulem}
\usepackage{color}
\usepackage{amsmath}
\usepackage{cite}
\usepackage[caption=false]{subfig}

 \usepackage{lipsum}

\usepackage{ragged2e}
\usepackage{sectsty}
\sectionfont{\fontsize{15}{15}\selectfont} 
\subsectionfont{\fontsize{13}{15}\selectfont}
\usepackage{tikz} 
\usetikzlibrary{shapes,arrows}
\usepackage[utf8]{inputenc}
\usepackage{authblk}

\usepackage{setspace}

 \newgeometry{a4paper, total={170mm,257mm},left=30mm,right=30mm,bottom=30mm,top=25mm,}

\begin{document}
\pagenumbering{gobble}
\title{\Large \textbf{Evaluating resilience in urban transportation systems for sustainability: A systems-based Bayesian network model}}
\centering
\author[1,2]{\small Junqing Tang\thanks{Email: jt746@cam.ac.uk}}
\author[2]{Hans Heinimann\thanks{Email: hans.heinimann@env.ethz.ch}}
\author[3]{Ke Han\thanks{Email: k.han@imperial.ac.uk}}
\author[4]{Hanbin Luo\thanks{Email: luohbcem@hust.edu.cn}}
\author[4]{Botao Zhong\thanks{Corresponding author. Email: dadizhong@hust.edu.cn}}

\affil[1]{\textit{Centre for Smart Infrastructure and Construction, Department of Engineering, University of Cambridge, Cambridge, CB2 1PZ, UK }}
\affil[2]{\textit{Department of Environmental Systems Science, ETH Zurich, Zurich, 138602, Switzerland }}
\affil[3]{\textit{ Department of Civil and Environmental Engineering, Imperial College London, London, SW7 2BU, UK}}
\affil[4]{\textit{School of Civil Engineering \& Mechanics, Huazhong University of Science \& Technology, Wuhan, 430074, China }}
\renewcommand\Authands{ }
\renewcommand\Authsep{  }
\renewcommand\Authfont{\bfseries}

\date{}

\maketitle

\begin{abstract}
\small This paper proposes a hierarchical Bayesian network model (BNM) to quantitatively evaluate the resilience of urban transportation infrastructure. Based on systemic thinkings and sustainability perspectives, we investigate the long-term resilience of the road transportation systems in four cities of China from 1998 to 2017, namely Beijing, Tianjin, Shanghai, and Chongqing, respectively. The model takes into account the factors involved in stages of design, construction, operation, management, and innovation of urban road transportation, which collected from multi-source data platforms. We test the model with the forward inference, sensitivity analysis, and backward inference. The result shows that the overall resilience of all four cities' transportation infrastructure is within a moderate range with values between 50\% to 60\%. Although they all have an ever-increasing economic level, Beijing and Tianjin demonstrate a clear ``V" shape in the long-term transportation resilience, which indicates a strong multi-dimensional, dynamic, and non-linear characteristic in resilience-economic coupling effect. Additionally, the results obtained from the sensitivity analysis and backward inference suggest that urban decision-makers should pay more attention to the capabilities of quick rebuilding and making changes to cope with future disturbance. As an exploratory study, this study clarifies the concepts of long-term multi-dimensional resilience and specific hazard-related resilience and provides an effective decision-support tool for stakeholders when building sustainable infrastructure.

\vspace{0.5cm}
\raggedright
\text{\textit{Keywords:}}
Resilience; Sustainable development; Transportation infrastructure; Bayesian networks; Systemic thinking.

 \end{abstract}

 \newgeometry{a4paper, total={170mm,257mm},left=35mm,right=35mm,bottom=30mm,top=30mm,}
\pagenumbering{arabic}
\cleardoublepage
\setlength\parindent{2em}\justify 

\section{Introduction}
As a type of fundamental infrastructure to ensure mobility and strength of the economy, resilient transportation systems are critical in the discussion of building smart and sustainable cities~\citep{faturechi2014measuring}. Indeed, transportation infrastructure is prone to a wide range of acute shocks such as terrorist attacks and natural disasters~\citep{reggiani2015transport}. For example, on 22 February 2011, a moment magnitude 6.2 earthquake happened in the city of Christchurch, New Zealand, caused 181 fatalities and massive infrastructure damage, including buildings and their road system~\citep{kaiser2012mw}. Additionally, chronic stresses such as severe congestion and inadequate accessibility incurred in network design, construction, operation, and management stages also play critical roles in affecting the efficiency and resilience of our urban transportation systems~\citep{ganin2017resilience,zhang2019scale}. To cope with these disturbances in long-term sustainable development, resilience has been paid increasingly more attention by urban decision-makers and practitioners.


Resilience is so multi-facet that it is sometimes used interchangeably with other similar concepts, or suffered from different interpretations in various fields~\citep{mattsson2015vulnerability}. It is not surprising that resilience, as an interdisciplinary concept, can be defined in many different ways~\citep{fisher2015disaster}. In urban sustainability studies, resilience has been largely understood as a design objective for buildings, infrastructure systems, communities, and planning and management~\citep{godschalk2003urban}. In systems engineering, resilience has been treated as a recursive process that includes cognitive and enabling functions such as anticipation, learning, and adaptation for complex infrastructure systems to perform a series of pro-active actions~\citep{park2013integrating,naser2018cognitive}. Modern urban smart systems, such as intelligent transportation systems, are built upon the abilities to comprehensively and effectively manage multiple system qualities, such as reliability, affordability, and maintainability~\citep{tang2019assessment}. While we only have ample knowledge, methods, and tools to manage certain isolated qualities such as reliability, we still lack the capability to manage multi-facet qualities such as resilience in achieving a more sustainable built environment. However, a hurdle in the front is the difficulty to effectively measure this concept before making pre-emptive moves in the planning and management processes.


Empirically, the resilience evaluation of transportation systems is still a rather problematic issue as it is more difficult than the vulnerability measures~\citep{reggiani2015transport}. The main discourse on transportation infrastructure resilience relies on the application of the percolation theory. Most of the previous studies were carried out based on network topology and mobility measurements. By mimicking vandalism or random failures through the so-called ``node-and-link-removal" simulations, topological robustness, or sometimes called network resilience, is evaluated through monitoring the performance of a chosen network measures~\citep{zhang2015assessing, bhatia2015network}. For example, Cantillo et al.~\citep{cantillo2019assessing} proposed a model for transportation network vulnerability assessments, which can identify critical links for the development of high impact disaster response operations. Wang et al.~\citep{wang2017multi} performed a substantial investigation on ten theoretical and four numerical robustness metrics and their performance through the robustness quantification of 33 metro networks under random failures and targeted attacks. A more realistic approach, with an emphasis on operational conditions, comes with edge weights, which indicate extra dimensions to the network topology such as travel time, cost, and travel distance~\citep{calvert2018methodology}. Also, the travel demand, route choice problems, and user equilibrium are taken into account as well~\citep{murray2004methodology, scott2006network}. However, studies in this mainstream have often overlooked the effect of long-term development and are mainly conducted from a unilateral assessment of the transportation systems. Furthermore, the coupling effect between resilient transportation development and associated urban factors such as regional economic development have relatively been less explored.

In systems engineering, transportation infrastructure is often treated as a functional system whose resilience is often perceived as an integrated system capability~\citep{boehm2017initial,bruneau2007exploring}. Systemic thinkings consider basic qualities in system resilience and have been applied in the transportation context. For instance, Wang et al.~\citep{wang2015day} proposed a day-to-day tolling scheme from a systems-based perspective to promote the rapidity of road traffic resilience in external disruptions. Tang et al.~\citep{tang2018resilience} proposed a congestion resilience metric based on the famous "4R" system functions for urban roads. Hosseini et al.~\citep{hosseini2016modeling} proposed a capability-based Bayesian network to measure system resilience in the case of a waterway port system. In addition, assessment methods and excellent reviews of resilience in generalized urban systems have also been pervasively documented, e.g.,~\citep{bhamra2011resilience, ouyang2012three, ouyang2012time, francis2014metric, parsons2016top, platt2016measuring,cere2017critical}. However, most of the research in this stream is tend to be context-dependent and event-specific. It is still in need of a systems-based approach to comprehensively understand the long-term multi-dimensional resilience in the transportation sector.

A similar discussion also can be found in engineering stage-related investigations, such as construction management community. Much effort has been dedicated to studying the resilience in terms of disaster risk reduction in built environment and infrastructure construction projects~\citep{bosher2011disaster}. Various discussion on the roles of project managers and stakeholders have been identified and reframed in terms of building disaster resilience into infrastructure management~\citep{dainty2008integrating,haigh2010integrative,crawford2013participatory}. For example, Bosher et al.~\citep{bosher2007integrating} used UK professionals from construction sectors as a case study and they reveal the insufficient links between construction-related stakeholders and resilient construction requirements. Wilkinson et al.~\citep{wilkinson2016improving} explored different approaches which the construction sector might improve its resilience during infrastructure projects. Nonetheless, an evaluation of resilience in transportation infrastructure that considers factors in the whole stages of design, construction, operation, management, and innovation has been rarely investigated.

Evaluating the resilience of transportation infrastructure is considered as a multi-attribute decision problem under a great amount of uncertainty and fuzziness. To bridge these aforementioned gaps, this paper constructs a novel hierarchical Bayesian network model (BNM) using a systems-based approach to capture the multi-dimensional consideration in measuring infrastructure resilience. The model covers stage-related indicators in the urban transportation sector from a sustainability perspective. Four cities from China are used as case studies to illustrate the effectiveness of the model and their long-term resilience-economic development are empirically studied. Insights and implications are also remarked concerning building resilience into urban transportation infrastructure, and a pathway to achieve sustainable development in the city built environment is also identified through the analysis. The main contributions of this paper can be summarized as follows:
\begin{itemize}
\item In this paper, we assess long-term multi-dimensional resilience of urban transportation systems, which provides a comprehensive evaluation based on a compounded consideration of the stages in design, construction, operation, management, and innovation. This bridges several major gaps in the literature and enriches the decision-support toolkit for policymakers, planners, practitioners, and other stakeholders.

\item The paper clarifies the concepts of long-term multi-dimensional resilience through systemic thinking and sheds new lights on the dynamic and time-varying characteristics of transportation resilience. Also, subsequent analysis reveals a pathway for building resilience into transportation systems, which could be useful to better understand resilience and sustainable development in urban transportation.

\item A high transferability of the proposed BNM could be expected due to the generality of the applied systems-based reasonings. The model and approach demonstrated in this paper could be useful for other urban systems and contribute a broader scientific understanding in managing infrastructure resilience.
\end{itemize}

The remainder of the paper is organized as follows: In the next section, we introduce the methodological background of the Bayesian network theory, including the forward inference, sensitivity analysis, and backward inference. Section 3 presents the reasonings and frameworks we used for model construction, including the hierarchical structure and dependence logic of the variable nodes. Section 4 demonstrates the case studies and empirical findings from the model inference. The coupling strength between transportation resilience development and regional economic development is also analyzed in this section. After that, Section 5 summarizes several practical implications regarding infrastructure resilience and sustainability and discusses the limitations of the study. Finally, we end the paper in Section 6 by concluding major findings and remarks.



\section{Methodology}
\subsection{Bayesian networks}
A Bayesian network model (BNM), also known as a Belief network model, is a direct acyclic graph (DAG). It is a combination of graph theory and probability theory. The idea is that, given a random variable X, another set of variables may exist that directly affect that variable X's value. Bayesian networks are widely recognized as an effective and developed technique, based on Bayes' Theorem, for tackling probabilistic assessments with multiple variables~\citep{wu2015dynamic}. It is prominent for multi-dimensional and multi-facet evaluation but with little application in resilience modeling~\citep{hosseini2016modeling}.

\begin{figure}[htb!]
\centering
\includegraphics[width=1\textwidth]{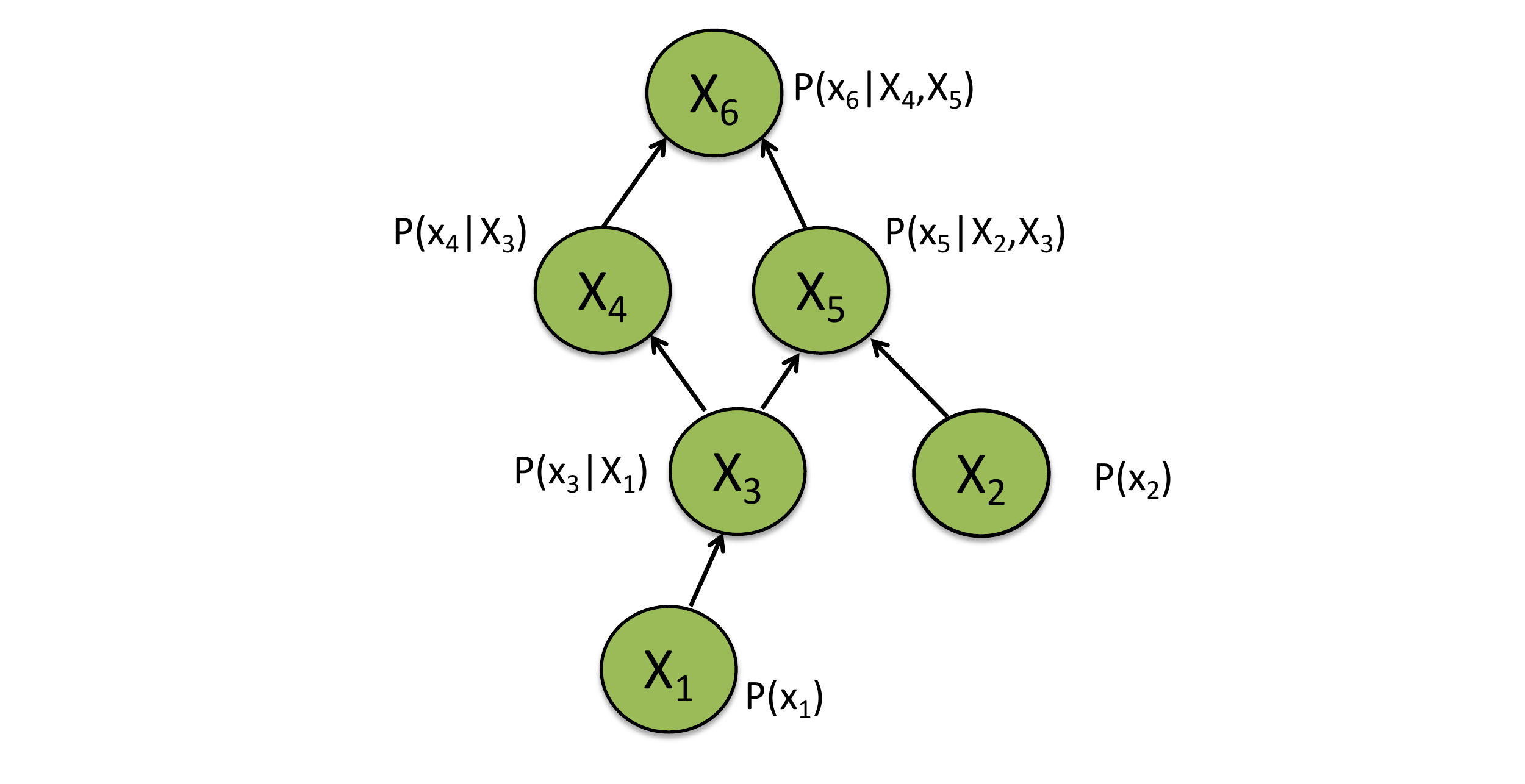}
\caption{An illustrative example of BNMs. Node $X_{1}$ and $X_{2}$ are root nodes, $X_{3}$, $X_{4}$ and $X_{5}$ are intermediate nodes and node $X_{6}$ is a leaf node, which is also the target node in calculation.}
\label{fig 1} 
\end{figure}

In a BNM, variables are abstracted as nodes and their conditional dependence is represented as edges/links, forming a directed network~\citep{heckerman2008tutorial}. The essence of BNMs is to compute the posterior probability distribution of target variables (or unobserved variables) conditioned on input variables (or observed variables). Mathematically, let $V=\{X_{1}, X_{2}...X_{n}\}$ be the variables in a BNM, where the conditional dependence among variables are represented by the topology of the network. Given that an outbound edge exists from node $X_{1}$ to $X_{2}$, this connection indicates that the probability states of $X_{2}$ are dependent on the outcomes of the node $X_{1}$. It is defined that $X_{1}$ is the \textit{parent node} of $X_{2}$ and $X_{2}$ is the \textit{child node}. There are three types of nodes can be found in a BNM, that is, (1) nodes without inbound edges (no parent nodes), named \textit{root nodes}, (2) nodes without outbound edges (no child nodes) are labeled as \textit{leaf nodes}, and (3) nodes with both inbound and outbound edges are called \textit{intermediate nodes}. An illustrative example of BNMs is demonstrated in Fig.~\ref{fig 1}.

The interdependent relationship among nodes is defined by a joint probability distribution~\citep{zhang2013decision,zhu2003application}. Each node in the network is associated with a Conditional Probability Table (CPT) that defines the probability of the node states under the conditions of its parent nodes~\citep{CICTP}. Therefore, the joint probability for the full BNM structure can be calculated, using the chain rule, into a factorized form with all the ``parents"~\citep{sun2015bayesian}. In this way, the full joint probability distribution of a BNM contains $n$ variables can be expressed as:

\begin{equation}
P(X_{1}, X_{2}...X_{n}) = \prod_{i=1}^{n} {(P(X_{i}|\psi_{i}))}
\end{equation}
where $P(X_{1}, X_{2}...X_{n})$ is the full joint probability distribution of the BNM and $\psi_{i}$ is the parents of node $X_{i}$.

The design of a BNM often involves two steps, namely (1) to determine the structure; and (2) to set CPTs for variables, respectively. Conventionally, there are three ways to determine the structure of the BNM and its associated CPTs~\citep{sun2015integrated}. The first method is to learn the structure and the CPTs using algorithms from a large amount of historical data. The advantage of this method is that it is based on objective data and mathematically rigorous algorithms. Therefore, the objectivity and rigor of this method are prominent~\citep{zhang2013decision}. In addition, because it is data-driven, the resultant BNM is usually highly accurate for predictive problems. However, three prominent shortcomings of this method can be realized: (1) it is often limited by the data availability and quality; (2) for fuzzy and uncertain concepts like resilience, this method is not realistic because there is no benchmark to calibrate and learn all the parameters; and (3) the structure obtained after applying data-driven structure learning is often difficult to understand~\citep{khakzad2011safety}. This leads to the second method, which uses expert prior knowledge to estimate the full model structure and the conditional probability between variable pairs. This method is considered effective for modeling concepts with uncertainty and fuzziness such as resilience~\citep{hosseini2016modeling}. In this paper, expert knowledge, professional assessment, and logic reasonings on well-defined resilience frameworks were used to determine the structure and CPTs in the proposed model. Finally, the third method is to combine the first two during the BNM design process as a compromised solution.

\subsection{Deductive reasoning}
An important feature of a BNM is its ability to perform deductive reasoning, also called forward analysis, for prediction of the leaf node $(T)$ under the condition of inputs from the root node variables $(X_{1}, X_{2}, ..., X_{n})$. This analysis aims to grasp a knowledge of the target leaf node given that the states of root nodes are known. To perform this analysis, the probability of states in each root node is input as evidence into the model. Based on pre-defined CPTs, the evaluation of the probability distribution of $T$, represented by $P(T = t)$, can be calculated by the following equation.

\begin{equation}
P(T = t) = P(T = t | X_{1} = x_{1}, ..., X_{n} = x_{n}) \times P(X_{1} = x_{1}, ..., X_{n} = x_{n})
\end{equation}
where, $t = {t_{1}, t_{2}, ..., t_{P}}$ is a range of $P$ states for the node $T$, and $P(T = t | X_{1} = x_{1}, ..., X_{n} = x_{n})$ denotes the conditional probability distribution of $T$. $P(X_{1} = x_{1}, ..., X_{n} = x_{n})$ represents the joint probability distribution of $X_{i}$.

\subsection{Sensitivity analysis}
Sensitivity analysis of a BNM serves as a vital stage to investigate the most influential variables when calculating the target $T$. It presents the individual contribution of each root node variables to the occurrence of a particular state in target $T$. Because it is an in-depth diagnosis of the influential power of each factor variable, it is particularly useful for decision-makers as it tells which factors should be paid extra attention to achieve a higher target level in their systems. Many indicators and methods are available to conduct sensitivity analysis in BNMs. In this paper, we adopt a sensitivity measure presented by Zhang et al.~\citep{zhang2013decision} - Sensitivity Index ($SI$) - to assess the variance of the probability distribution of target node $T$ at state $t$ with respect to the changes of variables $X_{i}$ in the model. By tuning on evidence input in root nodes, the $SI$ can be calculated as follows.

\begin{equation}
SI (X_{i}) = Max \{P(T=t |X_{i} = x_{i})\} - Min \{P(T=t |X_{i} = x_{i})\}
\end{equation}
where $i$ = 1, 2, 3, ..., n. The Sensitivity Index ($SI$) gives a rank list of the tested variables of the BNM in which those at the top are considered as the most influential and key factors.

\subsection{Abductive reasoning}
In the mainstream of evaluating infrastructure resilience, performance-based metrics (PMs) and topological network-based models (TNMs) are popular and particularly adequate for index-based evaluations. However, the capability of performing abductive reasoning, also called backward analysis, in Bayesian networks is unparalleled to those traditional methods~\citep{zhang2013decision}. The abductive reasoning aims to find out the posterior probability of root node variables if a certain probability of being in state $t$ is set as evidence in the target node $T$. Given the posterior probability of factor $X_{i}$ when $T=t$ is $P(X_{i} | T=t)$, it can be calculated as follows.

\begin{equation}
P(X_{i} | T=t) = \frac{P(X_{i}=x_{i}) \times P(T=t | X_{i} = x_{i})}{P(T=t)}
\end{equation}
where $i$ = 1, 2, 3, ..., n. Abductive reasoning is often performed in steps. It is very likely that the factor $X_{i}$ is a dominant cause to a certain level of resilience if the posterior probability of factor $X_{i}$ is close to 100\%. After the determination of the first factor, this factor can be set as an additional evidence for the diagnose in the next step. A clear pathway to achieve a certain level of resilience can be obtained in a backward fashion by repeating this process step-by-step. This unique analysis is a reliable way to draw a roadmap for decision-makers and practitioners to build targeted level of infrastructure resilience.

\section{Model construction}
\subsection{Hierarchical layout of the BNM}
In this section, we present our systemic-thinking approach in designing an appropriate BNM for evaluating infrastructure resilience in urban transportation systems. We proposed a hierarchical layout for the model which consists of three distinct layers, namely \textit{Function layer}, \textit{Quality layer}, and \textit{Factor layer}, respectively, following a ``macro-meso-micro" paradigm of logic. The sub-structure of \textit{Function layer} and \textit{Quality layer} are anchored in well-established frameworks. Sustainability perspectives are applied when selecting the variables in each layer and reasoning their causal relationships. 

\subsection{Function layer}
The purpose of the \textit{Function layer} is to establish the macro-level functions that directly contribute to the overall infrastructure resilience. At this level, the reasoning process is based on established frameworks and commonly-agreed definitions proposed in the resilience engineering field.

\begin{figure}[htb!]
        \centering
	\includegraphics[width=1\textwidth]{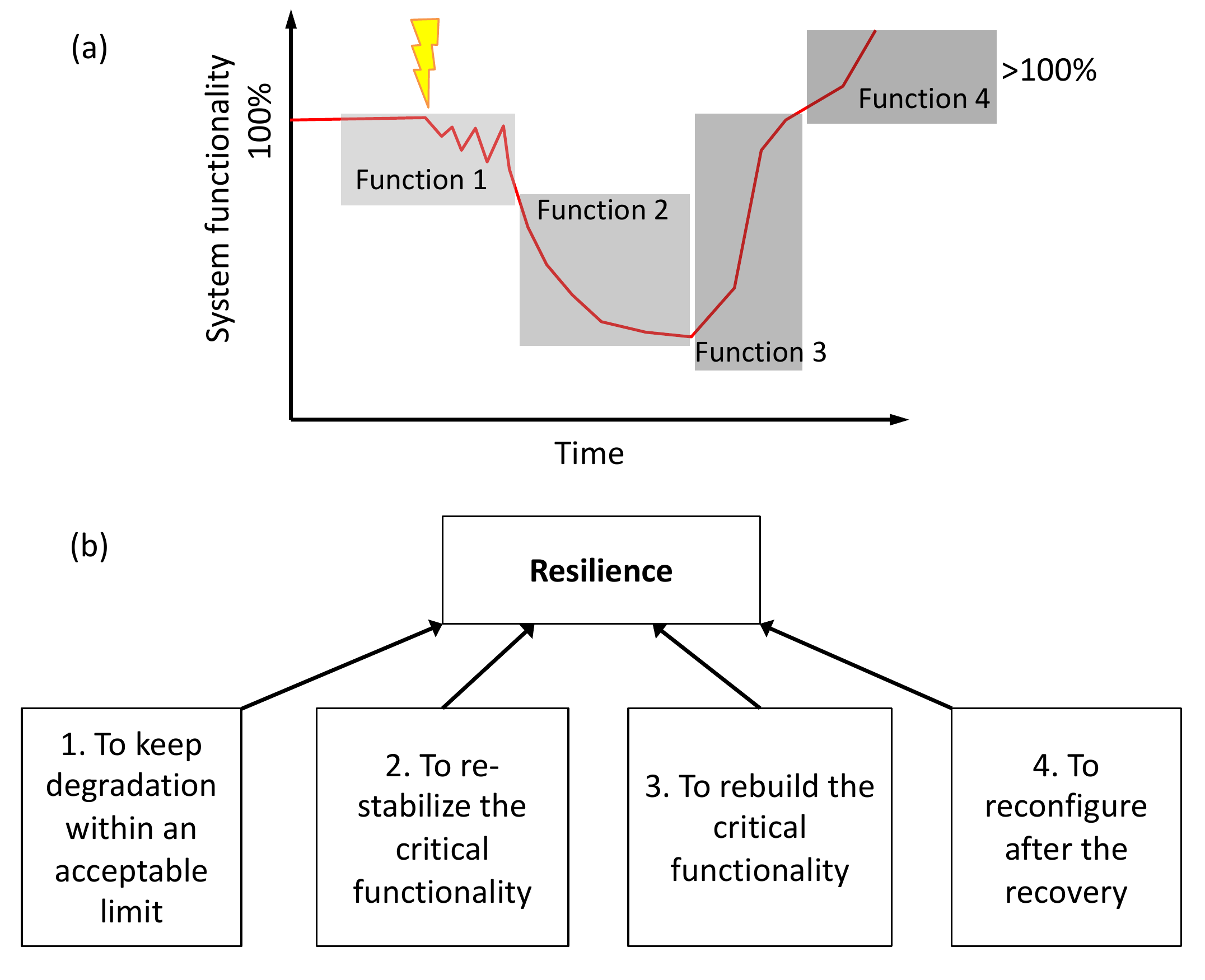}
	\caption{(a) Typical ``failure-recovery" performance with adaptive post-event recovery. (b) Reconstructed resilience framework with four transferable system functions. Adapted and modified based on Heinimann and Hatfield~\citep{HeinimannHatfield}.}
	\label{fig 2} 
\end{figure}

Heinimann and Hatfield~\citep{HeinimannHatfield} proposed a resilience framework in terms of transferable system functions, providing a comprehensive systems-based consideration (Fig~\ref{fig 2}). There, system resilience should include resistance, re-stabilization of critical functionality, the rebuilding, and reconfiguration of that functionality. These four functions provide a holistic characterization of the so-called ``failure-recovery" performance curve~\citep{tang2019assessment}, also known as resilience-triangle curve~\citep{bruneau2003framework} in a typical system resilience performance (Fig~\ref{fig 2} (a)). This particular shape is often triggered by an either external or internal disturbance, followed by immediate absorption and post-event reactions, including the abilities to restore and adapt~\citep{biringer2016critical}. From the perspective of socio-technical infrastructure systems, these four system functions are generally applicable across various urban systems (Fig~\ref{fig 2} (b)) and can be interpreted as follows.

\begin{itemize}
\item \textbf{To keep degradation within an acceptable limit}: This function indicates the infrastructure's ability to resist the negative effect of disturbance and maintain the level of functionality within an acceptable range~\citep{haimes2009complex}. This function is the key to sustainable development which stands for the fact whether the system can survive the impact without fatal collapse. The ``accepted limit" referred here is to emphasize whether the system's functionality has fault-tolerant and resistance attributes so that the impacts of the disturbance could be absorbed~\citep{henry2012generic}.

\item \textbf{To re-stabilize the critical functionality}: This function represents the ability of infrastructure systems to restore its stability. It answers the question: Can the system stabilize its level of functionality to prevent further unstable fluctuation? It can be seen as a connection between system resilience and fragility and plays an indispensable role in which absorptive and adaptive actions might take place~\citep{manyena2015bridging}.

\item \textbf{To rebuild the critical functionality}: The aim of this function is to quickly recover from the disturbance events. Once system performance has been stabilized and a total collapse has been successfully prevented, the recovery process can start to regain the critical functionality of the infrastructure systems. The realization of this function often requires the cooperation between social and technical efforts, such as proposing a rescue plan, organizing a rebuilding team, engaging with communities and residents, and managing special resources~\citep{linkov2013resilience,miao2013embedding}.

\item \textbf{To reconfigure after the recovery}: A resilient system should have the ability to perform reconfigurable recovery, which is manifested as a growing post-event functionality level that excels the pre-event functionality~\citep{tang2018resilience}. In other words, after recovering its basic functionality, it is important to think what can be learned from the event so that the design, construction, operation, and management can be re-organized and updated to make the system adaptive and more resilient for future events. The reconfiguration function in the framework is crucial for long-term resilient development as a proper configuration pattern may suggest a tendency towards a resilient or a vulnerable transportation network after disturbance events~\citep{reggiani2015transport,dehghanian2017quantifying}.

\end{itemize}

\subsection{Quality layer}
Ten fundamental system qualities are introduced to form the \textit{Quality layer} in the proposed BNM, including ``Availability", ``Changeability", ``Reliability", ``Maintainability", ``Serviceability", ``Robustness", ``Safety", ``Reparability", ``Affordability", and ``Adaptability". These ``-ilities" are widely found in infrastructure systems and act as mesoscopic enablers which contribute to the aforementioned four functions. The causal relationships among them were adapted from an ontology study conducted by Boehm et al.~\citep{boehm2013tradespace,boehm2017system} in a four-year exploratory project regarding trade space in the system's fundamental qualities. The project aimed to tackle the misuse problem of these system qualities. Some of these system qualities are sometimes used interchangeably and causing misunderstandings and confusion. For example, resilience is sometimes interpreted as robustness or reliability. This research pioneered the investigation of the system's qualities to decode their ontological relationships. Based on their framework, we improved the ontological structure so that it is suitable for infrastructure systems. 

\begin{figure}[htb!]
        \centering
	\includegraphics[width=1\textwidth]{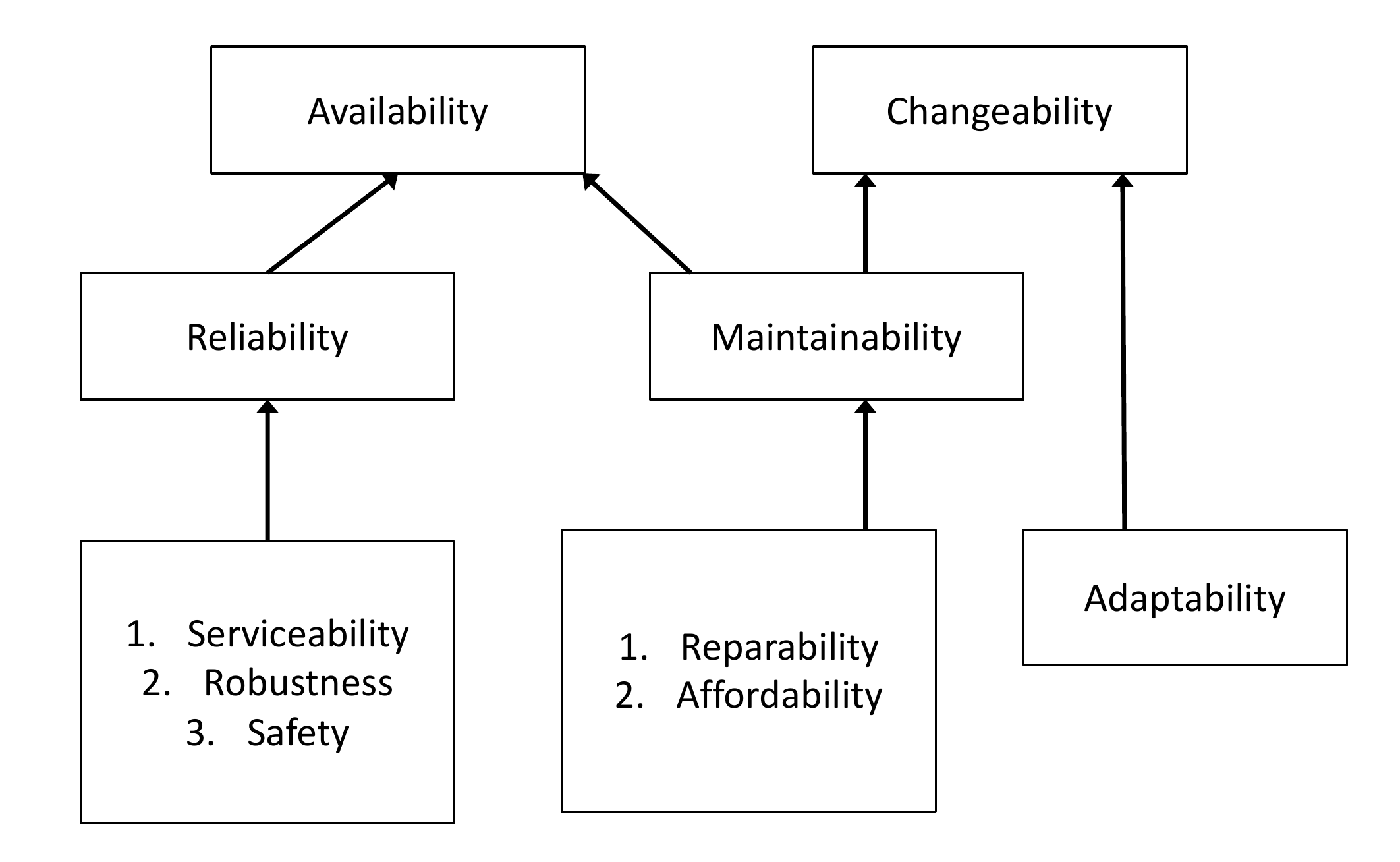}
	\caption{Modified Boehm's ilities ontology structure}
	\label{fig 3} 
\end{figure} 

The overall structure of the quality layer is shown in Fig~\ref{fig 3}. System resilience, as a system property, is argued as a compound attribute that formed from two qualities, namely ``Availability" and ``Changeability"~\citep{boehm2017system}. As well defined in reliability engineering domain, ``Availability" is clearly in the relation between ``Reliability" and ``Maintainability" in terms of Mean Time to Failure (MTTF) and Mean Time to Repair (MTTR). On the other hand, a system's ``Changeability" should be defined by the capability to adapt based on the current state (``Maintainability" and ``Adaptability"). After that, it only takes a second to realize that ``Reparability" and ``Affordability" should have a direct causal feedback loop to ``Maintainability" as ``Reparability" sustains the long-term functionality and ``Affordability" indicates that the system has to be economically sustainable to be maintained~\citep{boehm2017initial,boehm2016key}.

Similar to the generic functions mentioned before, these qualities are widely found in most critical infrastructure systems. In the transportation context, the interpretation of these qualities can also be realized. For example, the overall level of service (in terms of passenger traffic and overall congestion) and safety (in terms of traffic accidents) of a transportation system has a profound influence to sustain its reliability. Besides, the system's capability to adapt new technologies and knowledge through innovation would mainly contribute to its long-term progress and the changeability. Eventually, from a wide-angle view, the integrated overall effects of all these qualities ensure a resilient development in the transportation systems.




\subsection{Factor layer}
\textit{Factor layer} acts as a micro-level bottom in the proposed BNM, which enables the application of the quality layer. Although much research has been dedicated to the measurement of infrastructure resilience, it remains an intangible term with great fuzziness. As a consequence, no universal rule exists in the choice of an appropriate set of indicators for system qualities when modeling infrastructure resilience. For transportation systems, we selected 14 factors to form a \textit{Factor layer} as root nodes in the model. During the selection, we considered factors involved in the design, construction, operation, maintenance, management, and innovation in the transportation sector to ensure the comprehensiveness and universality. Furthermore, to avoid possible colinearity caused by excess variables, we only selected the most representative factors. The fators are allocated to nodes in the \textit{Quality layer} based on their natural attributes. For example, the number of annual traffic accidents could be used to denote safety and road density in a city's urban area could be selected to indicate serviceability. One thing noteworthy is that we use both quantity and quality of the technological and research innovation (in terms of the numbers of submitted patents and granted patents) to denote system’s adaptability for change and improvements~\citep{lanjouw2004patent, bettencourt2007growth,wu2007use}. Table 1 illustrates a summary of the variables in all three layers and the attributes of the selected factors. 

\begin{table}[htb!]
\label{tab 1}
\caption{A summary of all the variables in the proposed BNM}
\begin{tabular}{ll}
\hline
Label & \textbf{Function layer} \\ \hline
X1 & To keep degradation within an acceptable limit \\
X2 & To re-stabilize the critical functionality \\
X3 & To rebuild the critical functionality \\
X4 & To reconfigure after the recovery \\ \hline
 & \textbf{Quality layer} \\ \hline
X5 & Availability \\
X6 & Changeablity \\
X7 & Reliability \\
X8 & Maintainability \\
X9 & Serviceability \\
X10 & Robustness \\
X11 & Safety \\
X12 & Reparability \\
X13 & Affordability \\
X14 & Adaptability \\ \hline
 & \textbf{Factor layer} \\ \hline
S1 & Road density in urban area (\%) \\
S2 & Per capita area of paved roads (sq.m$^*$) \\
S3 & Congestion level \\
S4 & Passenger traffic of urban roads (10000 persons) \\
R1 & Direct economic loss affected by natural disasters (100 million Yuan) \\
R2 & Investment of projects in prevention of disasters (10000 Yuan) \\
SA1 & Number of injuries on traffic accidents (person) \\
SA2 & Number of traffic accidents (Time) \\
RE1 & Net growth of paved roads (sq.m) \\
RE2 & Number of employed persons in road transport (person) \\
AF1 & Ratio of affordability Per capita of household on urban transport \\
AF2 & Ratio of affordability of local governments on urban transport \\
AD1 & Number of granted patents in transport research \\
AD2 & Ratio of granted patents in transport sector \\ \hline
\end{tabular}

* \small{sq.m - square meter; Yuan - Chinese currency; Congestion level and all the ratios are dimensionless.}
\end{table}

\subsection{Baseline of the BN model}
\begin{figure}[htb!]
        \centering
	\includegraphics[width=1\textwidth]{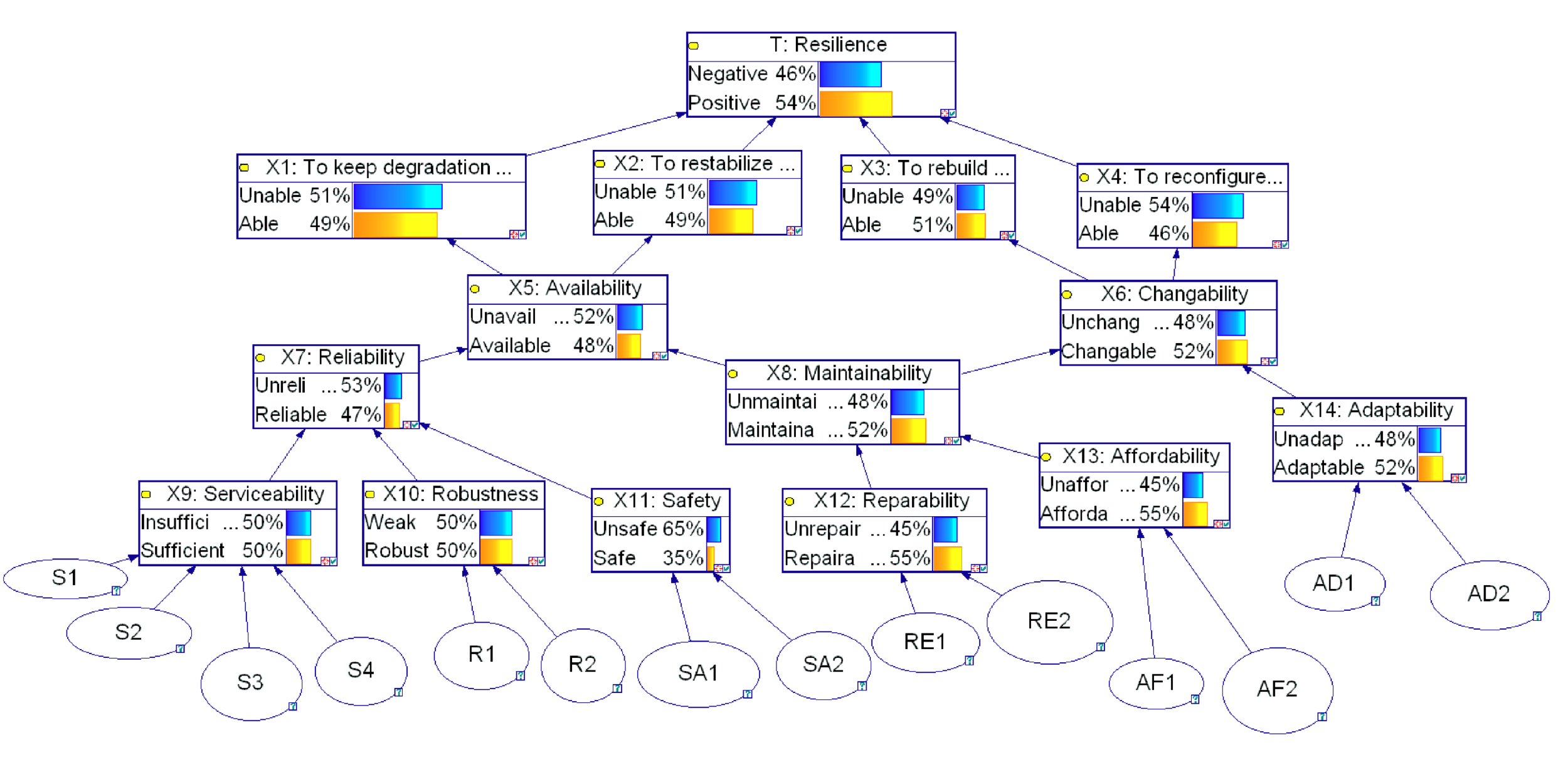}
	\caption{Baseline of the proposed BNM. The leaf node ``Resilience" is labeled as target $T$.}
	\label{fig 4} 
\end{figure}

With the layers being established, we next determine the category of the variables. Apart from the factor $S_{3}$ (Congestion level), the target node $T: Resilience$ and other variables in \textit{Function, Quality} and \textit{Factor layers} are set to be boolean variables, which has two states as ``Positive" and ``Negative" or ``True" and ``False". For example, node $T: Resilience$ has ``Positive" and ``Negative" and node $X_{7}: Reliability$ has ``Unreliable" and ``Reliable". The factor $S_{3}$ is set to be a categorical variable with three outputs on ``1" or ``0", namely ``Low: $<$ 1.5", ``Medium: 1.5 - 1.8", and ``High: $>$ 1.8", respectively. With all the variables being determined, we can obtain the structure of BNM as shown in Fig~\ref{fig 4}.

\begin{figure}[htb!]
        \centering
	\includegraphics[width=1\textwidth]{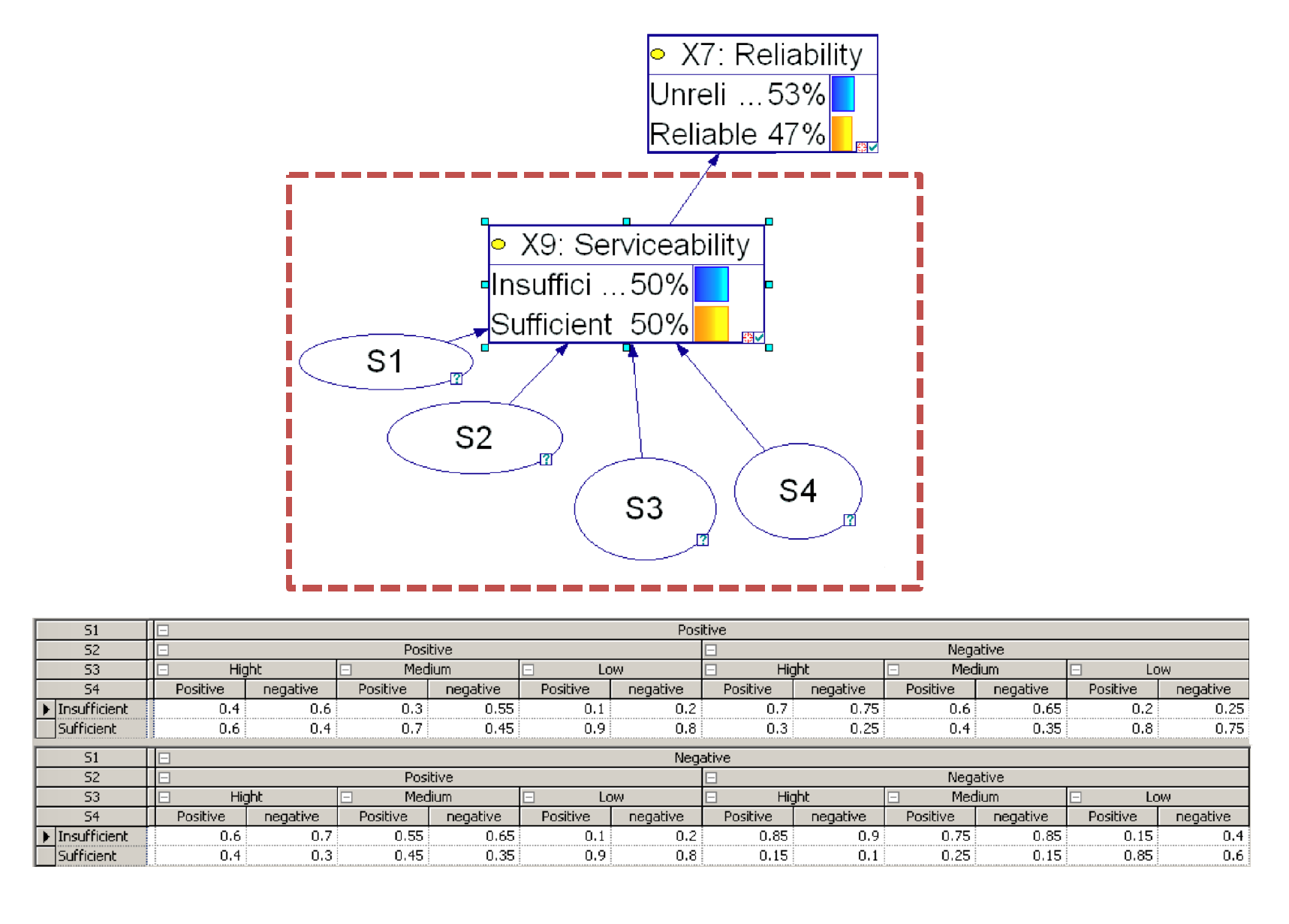}
	\caption{The CPT determined in $X_{9}$ - Serviceability}
	\label{fig 4_2} 
\end{figure}

Fig~\ref{fig 4_2} shows the determined CPT for variable $X_{9}$ - Serviceability. Note that the factor $S_{3}$ has three states and the other three factors have binomial states, which result in 24 scenarios to assess. In forward inference, the probability of $X_{9}$ is calculated using all the inputs from $S_{1-4}$ based on this conditional probability rule. The CPTs in other variables are determined in a similar way. In this way, we finally obtain a baseline model for the proposed BNM.

\section{Case study}
\subsection{Data description}
We selected four municipalities in China as the case study, namely Beijing, Tianjin, Shanghai, and Chongqing. These four cities are ranked as top 10 in China Top 100 Cities List published by Warton Economic Institute~\citep{100Cities}, which presents a top-level of China's urban economic development in recent years. We collected open-source statistical data for factors in Table 1 from 1998 to 2017. They were collected from multiple sources, including the China National Bureau of Statistics~\citep{CNBS}, China Statistical Yearbook~\citep{Yearbook}, United Nations Statistics Division~\citep{UN}, and Alibaba's AutoNavi Annual Reports~\citep{Alibaba}. The row dataset contains missing data and they were treated with the following methods: (1) For those indicators with no apparent growth rate and a small number of missing blanks, we fill in the blanks with the median value; And (2) For those with a linear or exponential growth, we fill in the missing data with regression predictions. Note that factors such as $S_{2}$ and $AD_{2}$ were calculated from the raw data. For example, $S_{2}$ can be calculated as the area of paved roads divided by the population in the urban area.

\subsection{Probability data assessment}
\begin{figure}[htb!]
        \centering
	\includegraphics[width=0.8\textwidth]{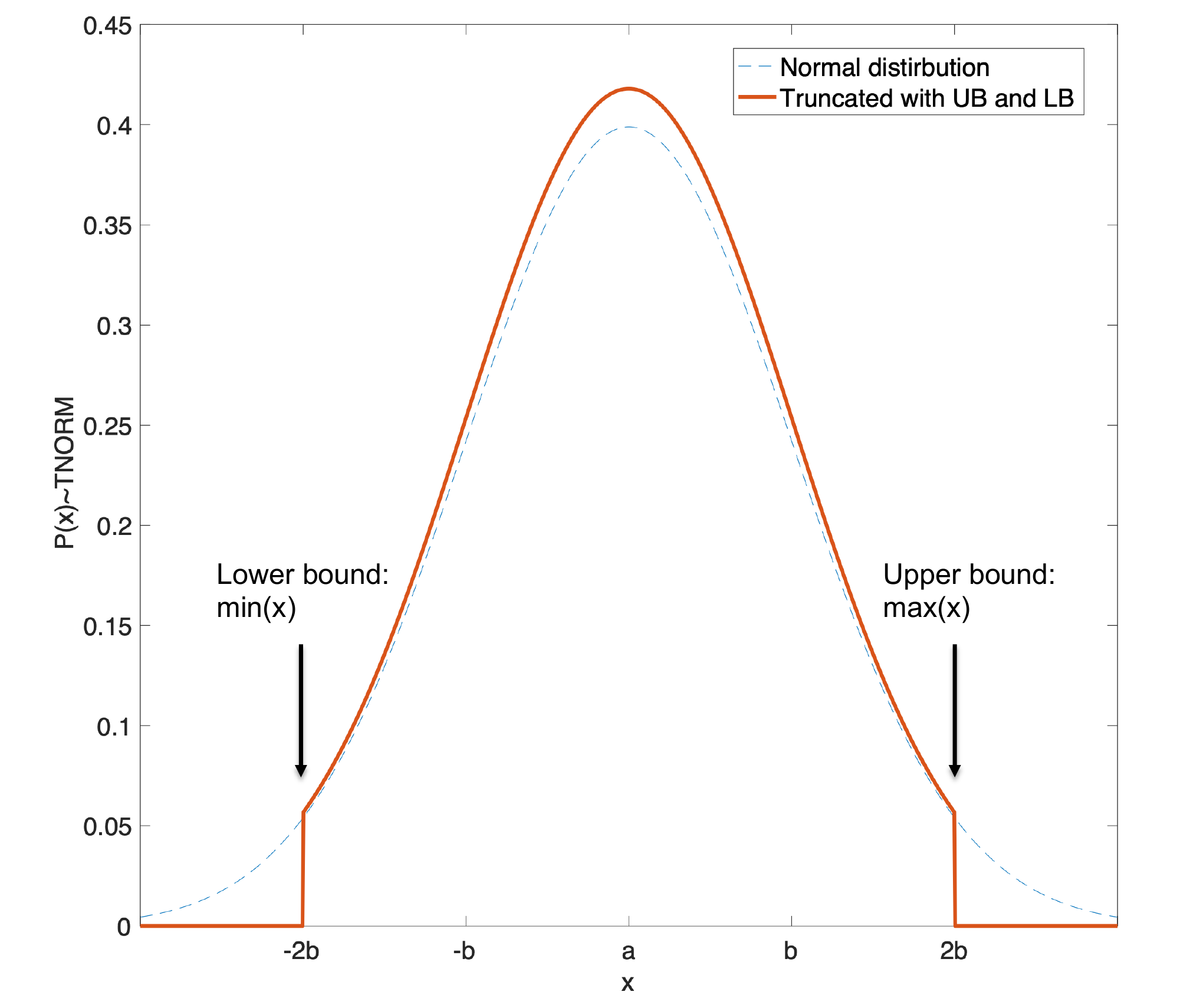}
	\caption{An illustrative example of a truncated normal distribution with $\mu$ = a, $STD$ = b, $LB$ = $min(x)$, and $UB$ = $max(x)$. Note that this is an illustrative case for TNORM as the $min(x)$ and $max(x)$ of factors may not exactly lie at -2$STD$ and 2$STD$, respectively.}
	\label{fig 4_3} 
\end{figure}

The historical statistical data have to be converted to prior probability distribution as inputs to the proposed BNM. For discrete variables such as $S_{3}$, it is straightforward to convert index values into categorical readings and then input as ``0" or ``1" in corresponding categories. However, for continuous variables such as $S_{4}$ - Passenger traffic, we used a truncated normal distribution (Fig~\ref{fig 4_3}) to model the prior probability distribution~\citep{hosseini2016modeling}. Generally, the truncated normal distribution is an appropriate method because it can be confined between determined lower and upper bounds, especially for modeling variables such as passenger traffic or other similar variables. For example, a truncated normal distribution denoted as TNORM with a mean ($\mu$) of $a$, standard deviance ($STD$) of $b$, upper bound ($UB$) of $max(x)$, and lower bound ($LB$) of $min(x)$ was applied to model the prior probability input for factor $S_{4}$.

\begin{equation}
S_{4}: (fit \sim TNORM) \sim 
(\mu = a, STD = b, LB = min(x), UB = max(x), 1)
\end{equation}

In this case, $\mu$ was calculated as 34596 for all four cities with $STD$ = 40840, $UB$ = 2767722, and $LB$ = 1127. ``1" denotes that cumulative probability was calculated as the inputs. In this way, the initial inputs of the prior probability for each factor can be modeled accordingly and the resilience evaluation analysis can be carried out in the next step. Table 2 demonstrates the input evidence in factor $S_{1}$ and $S_{2}$ for all four cities.

\begin{table}[htb!]
 \centering
\label{tab 2}
\caption{Input evidence in factor $S_{1}$ and $S_{2}$ as illustrative examples}
\begin{tabular}{llllll}
\hline
S1=positive & Year & Beijing & Tianjin & Shanghai & Chongqing \\ \hline
$\mu$ = 0.107 & 1998 & 1.30\% & 7.70\% & 47.28\% & 36.98\% \\
STD = 0.043 & 1999 & 1.90\% & 14.87\% & 46.00\% & 38.38\% \\
UB = 0.255 & 2000 & 2.68\% & 24.12\% & 44.84\% & 39.83\% \\
LB = 0.012 & 2001 & 3.69\% & 34.32\% & 43.79\% & 41.32\% \\
 & 2002 & 4.95\% & 44.36\% & 42.84\% & 42.85\% \\
 & 2003 & 6.49\% & 53.42\% & 41.98\% & 44.41\% \\
 & 2004 & 38.91\% & 60.14\% & 99.97\% & 44.76\% \\
 & 2005 & 74.54\% & 67.04\% & 99.97\% & 40.06\% \\
 & 2006 & 25.43\% & 82.46\% & 99.96\% & 50.47\% \\
 & 2007 & 13.67\% & 75.07\% & 38.49\% & 59.26\% \\
 & 2008 & 18.30\% & 80.30\% & 39.61\% & 60.17\% \\
 & 2009 & 18.18\% & 67.38\% & 35.80\% & 56.86\% \\
 & 2010 & 21.29\% & 73.10\% & 37.33\% & 56.69\% \\
 & 2011 & 22.42\% & 82.85\% & 38.95\% & 48.22\% \\
 & 2012 & 50.18\% & 89.57\% & 41.08\% & 56.05\% \\
 & 2013 & 49.42\% & 91.79\% & 43.04\% & 56.65\% \\
 & 2014 & 43.40\% & 91.20\% & 43.34\% & 60.21\% \\
 & 2015 & 45.50\% & 88.52\% & 46.60\% & 63.18\% \\
 & 2016 & 44.33\% & 80.33\% & 49.06\% & 71.74\% \\
 & 2017 & 40.42\% & 74.75\% & 51.99\% & 73.33\% \\ \hline
S2=positive & Year & Beijing & Tianjin & Shanghai & Chongqing \\ \hline
$\mu$ = 8.500 & 1998 & 11.31\% & 53.36\% & 15.32\% & 21.97\% \\
STD = 4.123 & 1999 & 12.42\% & 56.82\% & 15.16\% & 23.93\% \\
UB = 18.740 & 2000 & 13.60\% & 60.37\% & 14.62\% & 26.09\% \\
LB = 3.510 & 2001 & 14.86\% & 63.97\% & 14.62\% & 28.46\% \\
 & 2002 & 16.19\% & 67.60\% & 14.81\% & 31.06\% \\
 & 2003 & 17.60\% & 71.20\% & 14.88\% & 33.89\% \\
 & 2004 & 59.12\% & 57.98\% & 95.19\% & 28.85\% \\
 & 2005 & 69.05\% & 67.84\% & 78.69\% & 32.60\% \\
 & 2006 & 39.49\% & 90.81\% & 79.11\% & 46.53\% \\
 & 2007 & 24.10\% & 79.80\% & 16.60\% & 56.36\% \\
 & 2008 & 28.94\% & 92.34\% & 17.40\% & 59.49\% \\
 & 2009 & 28.44\% & 89.90\% & 16.48\% & 62.19\% \\
 & 2010 & 23.87\% & 93.94\% & 13.97\% & 58.36\% \\
 & 2011 & 21.60\% & 98.09\% & 13.97\% & 68.02\% \\
 & 2012 & 41.08\% & 98.85\% & 14.19\% & 70.07\% \\
 & 2013 & 41.46\% & 99.35\% & 14.35\% & 74.61\% \\
 & 2014 & 39.86\% & 97.68\% & 14.35\% & 77.97\% \\
 & 2015 & 41.55\% & 96.59\% & 15.25\% & 80.54\% \\
 & 2016 & 41.55\% & 95.27\% & 15.83\% & 81.72\% \\
 & 2017 & 39.86\% & 98.47\% & 16.66\% & 84.41\% \\ \hline
\end{tabular}
\end{table}

\subsection{Resilience analysis}
\subsubsection{Forward inference}
The forward inference aims to calculate the probability of the transportation systems being overall resilient. In this phase, decision-makers do not have comprehensive understandings about the system's resilience performance. With available historical data of all the factors, it is essential to calculate the resilience probability using forward inference in BNM. The prior probability of factors is set as input evidence for each year in each city case study. The probability of $Resilience=Positive$ is obtained as an output.

\begin{figure}[htb!]
        \centering
	\includegraphics[width=1\textwidth]{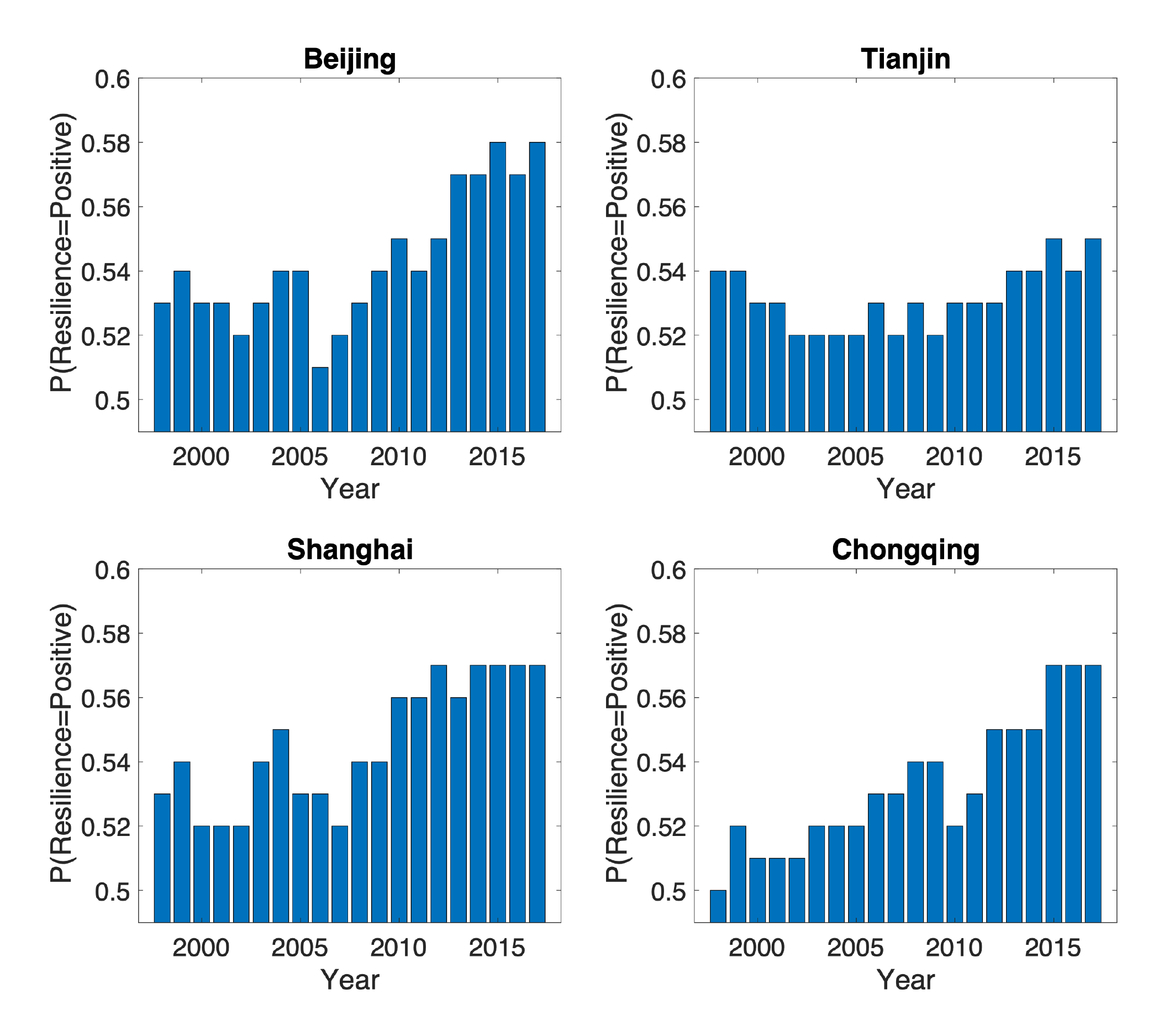}
	\caption{Calculated probability of resilience level for four cities of China from 1998 to 2017.}
	\label{fig 5} 
\end{figure}

Fig.~\ref{fig 5} shows the quantification results for all four cities from 1998 to 2017. A prominent feature is that the transportation infrastructure systems in these four cities have a moderate level of being overall resilient, from 50\% to 58\%. The maximum value (58\%) is found in Beijing and the minimum level of being resilient (50\%) is from Chongqing. However, the overall values from 2015 to 2017 are higher than that of the most past years in all four cities, which demonstrate a growing resilience in their transportation systems. Of interest is the temporal patterns of transportation resilience. Beijing and Tianjin demonstrate a rough ``V" shape trend in the profile during the 20 years of development, while Shanghai and Chongqing have a relatively more steady increase over time. Such patterns in Beijing and Tianjin could indicate that the overall resilience of the urban transportation system has a dynamic character in long-term development. In addition, it is clear that Beijing and Shanghai demonstrate a higher increasing rate in transportation resilience from 2005 to 2017. In contrast, Chongqing has a more steady increasing rate during the entire 20 years, while Tianjin has the lowest increasing rate in these four cities.

\begin{figure}[htb!]
        \centering
	\includegraphics[width=1\textwidth]{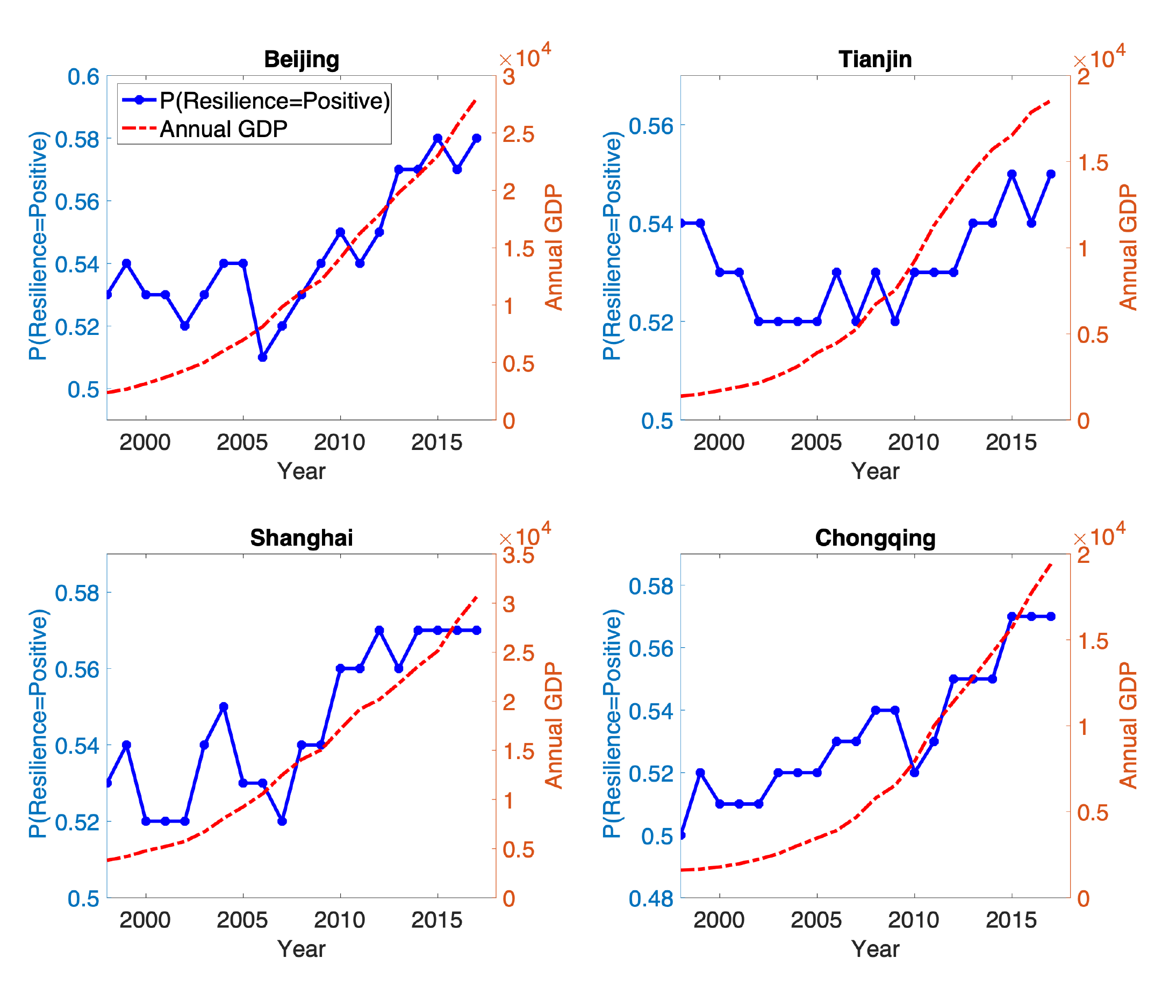}
	\caption{Resilience vs GDP}
	\label{fig 6} 
\end{figure}

\begin{figure}[htb!]
        \centering
	\includegraphics[width=1\textwidth]{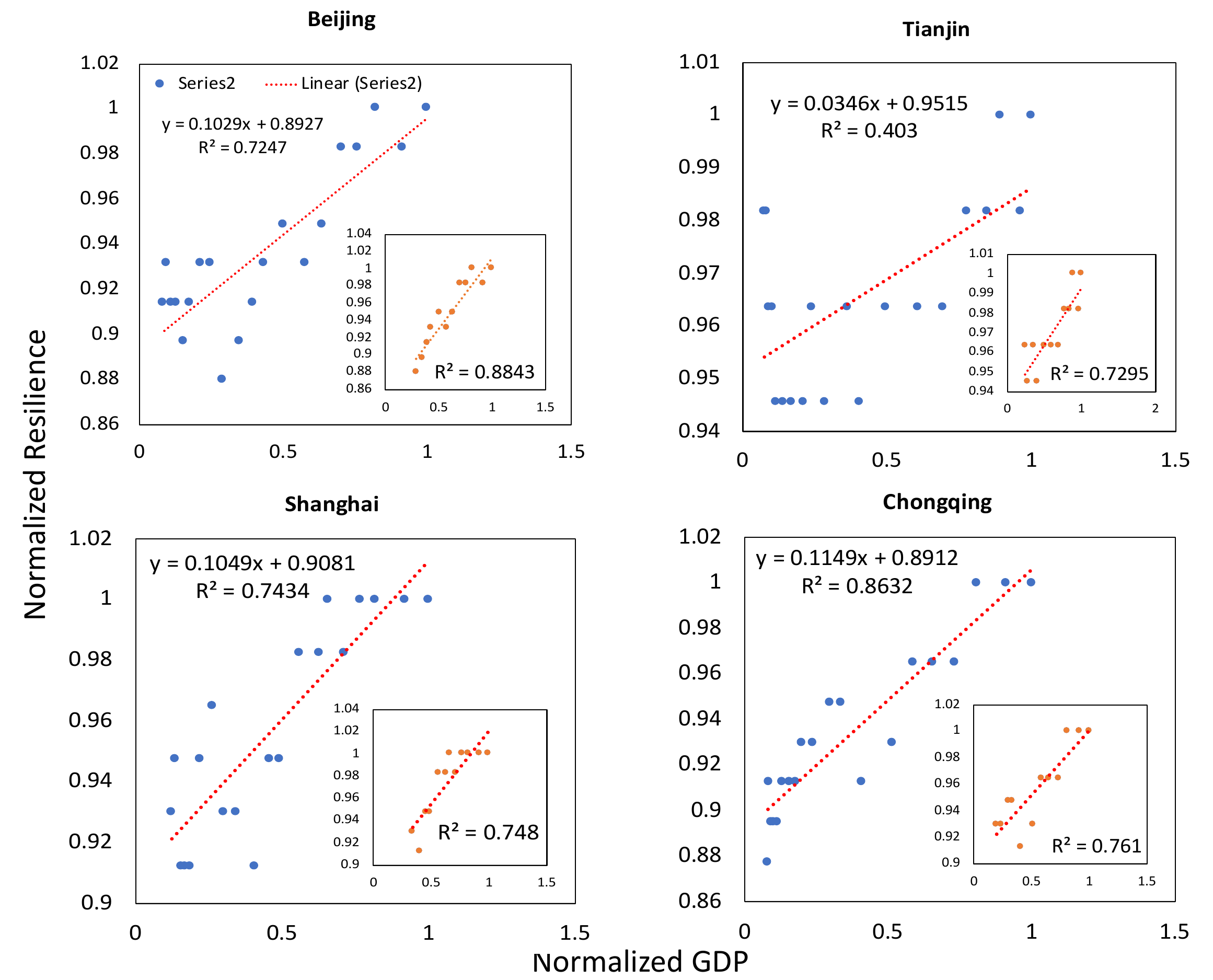}
	\caption{Resilience-Economic coupling effect. Note that the subplots are segment from 2005 to 2017.}
	\label{fig 6_2} 
\end{figure}

Comparing with the city's annual GDP development profile (Fig~\ref{fig 6}), we found that the long-term growth of resilience does not always correlate with economic growth at all time. Apart from Chongqing, the resilience values of the other three cities show a positive correlation with GDP after the year 2005. However, before 2005, there are no such positive correlations. In other words, while these major Chinese metropolia witness an outstanding growth in their economic development, it does not necessarily lead to an ever-increasing level of urban transportation resilience, which indicates a non-linear coupling effect between these two, especially before and after at the turning point of 2005 (see Tianjin and Beijing cases). This is also confirmed by the following resilience-economic coupling analysis (Fig~\ref{fig 6_2}). The regression and the $R^2$ indicates the coupling strength of the two variables. The subplots are the segmented analysis from 2005 to 2017 in which a better $R^2$ value indicates a positive coupling effect between transportation resilience and regional economic level (except Chongqing, there is a decrease in $R^2$ in the subplot). Particularly, the non-linear characteristic is relatively more obvious in Tianjin case, where the $R^2$ for overall fitting is around 0.4. However, this value promptly increases to 0.7 from 2005 to 2017.

Conventionally, it is considered that the economic growth would promote positive improvements in infrastructure and therefore enhance its overall resilience performance, such as more investments and better road networks. However, by seeing resilience as a multi-attribute system property, the positive economic development might bring up new urban mobility issues such as massive congestion and safety concerns, which could deteriorate the overall resilience of the transportation systems in our cities. This counter-intuitive finding could indicate that the level of economic maturity is one important index but it certainly not the only dominant role when evaluating long-term transportation resilience. It could also remind our city governors that transportation resilience might not always be enhanced merely by a one-sided emphasis on economic development and investment. Therefore, a co-evolved strategy in multiple aspects might be more feasible for building resilient transportation systems for long-term sustainable development.

\subsubsection{Sensitivity analysis}

The sensitivity test can also be viewed as the importance analysis. It aims to identify key factors/variables which have a greater influence on the evaluation results. With such identifications, decision-makers can pay extra attention to these checkpoints during the decision-making process. For knowledge-based Bayesian models, conducting sensitivity tests could also act as an alternative way to perform model validation~\citep{hanninen2014bayesian,hosseini2016general}. Using Eq.3, we are able to calculate the sensitivity measure index $SI$ for each variable in the proposed BNM.

\begin{figure}[htb!]
        \centering
	\includegraphics[width=1\textwidth]{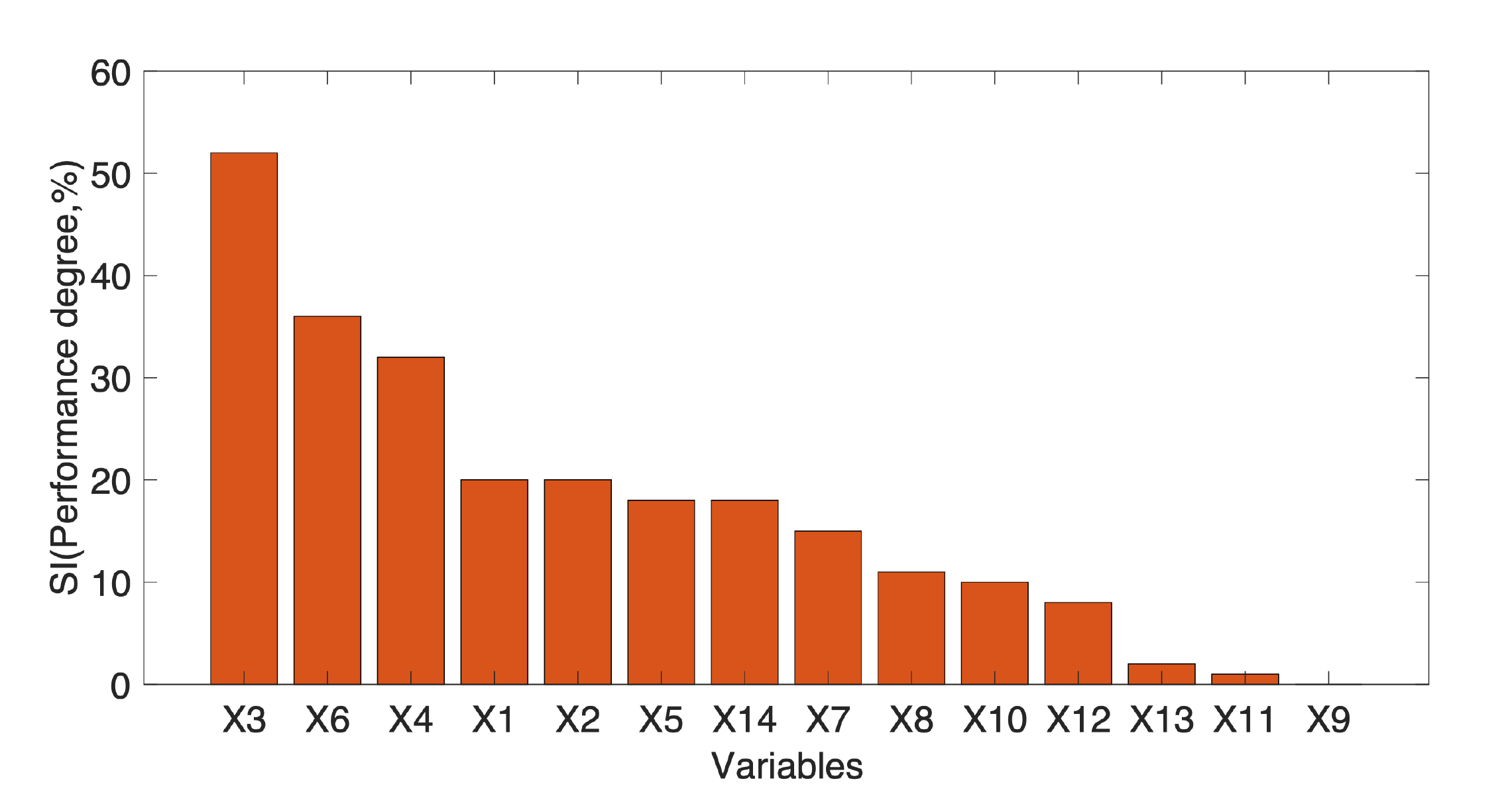}
	\caption{Sensitivity test score SI for variables}
	\label{fig 7} 
\end{figure}

As shown in Fig.~\ref{fig 7}, we rank the sensitivity scores of each variable in descending order. Variable $X_{3}$  (To rebuild the critical functionality), $X_{6}$ (Changeability), and $X_{4}$ (To reconfigure after the recovery) are the top three of the ranking list, which indicates that the overall transportation resilience is remarkably sensitive to these three variables. Meanwhile, it also indicates that the abilities to quickly rebuild and to make changes by reconfiguration in order to cope and adapt the post-event phases in a city's transportation system play indispensable roles in the determination of its system resilience. This result is in line with our mutual perception on infrastructure resilience that a resilient transportation system should have a quick ``bounce back" and certain flexibility for future adaptation. Therefore, it is of importance to place more emphasis on these three variables when managing the transportation infrastructure.

\subsubsection{Backward inference}
One of the prominent merits of the Bayesian networks is that it can perform abductive reasonings about the evaluation. Compared with traditional resilience assessment methods such as deterministic metrics and performance-based tools, BNM is able to tell more about the posterior probability of each variable when certain probability level is observed or set as the target in the leaf node. This particular trait allows decision-makers to perform in-depth diagnosis on system performance in a backward fashion with a clear target level of resilience which they would like to ultimately achieve. By doing so, we are able to model the evolution pathway of reaching a certain level of resilience in the transportation systems.

\begin{figure}[htb!]
        \centering
	\includegraphics[width=1\textwidth]{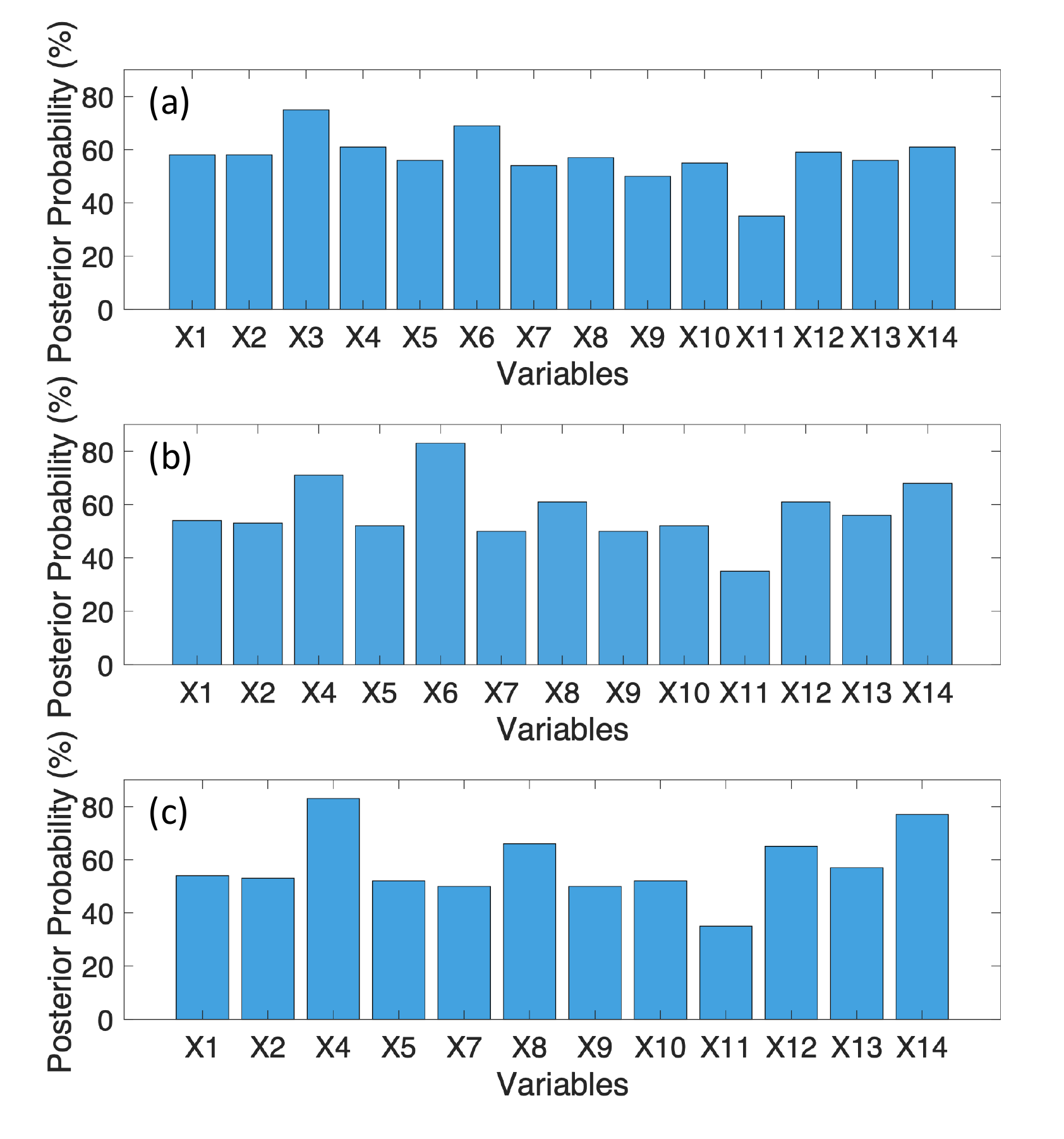}
	\caption{Backward diagnosis of transportation infrastructure resilience. (a) $P(X_{i}|T=Positive)$; (b) $P(X_{i}|T=Positive, X_{3}=Able)$; and (c) $P(X_{i}|T=Positive, X_{3}=Able, X_{6}=Changable)$}
	\label{fig 8} 
\end{figure}

Fig~\ref{fig 8} demonstrates the calculated posterior probability distribution of the variables $X_{1}$ to $X_{14}$ with a target level of resilience at 100\% positive. In subplot (a), it is clear that $X_{3}=Able$ (with a probability of over 60\%) is most likely to be the ``driving force" for achieving such resilience level. In the next step, $X_{3}=Able$ can be set as a piece of extra evidence to the model to find out the second most-likely cause. Fig~\ref{fig 8} (b) shows that $X_{6}=Changeable$ has the highest posterior probability in this step, which should be the focal and prioritized variable at this step. The result (both $X_{3}$ and $X_{6}$ being set as evidence) shown in subplot (c) indicates that the third variable is $X_{4}$ with a probability around 80\%. For the next step, all these three variables can be set as extra evidence to find the subsequent target variable. In this way, a pathway to achieve a high resilience level in urban transportation systems can be identified in a step-by-step manner.

\section{Implications and limitations}

According to the World Economic Forum, the world is at the beginning of a fourth industrial revolution due to the emerging advanced technologies, which would involve discourse about digital and smart infrastructure~\citep{WEF}. One of the advantages of building smart infrastructure assets is that it will allow shareholders to get more on increasing capacity, efficiency, reliability and resilience~\citep{bowers2017smart}. On the other hand, according to United Nations' Sustainable Development Goals, building resilient infrastructure, promoting inclusive and sustainable industrialization and fostering innovation have been identified as key components to support economic development and human well-being~\citep{sciamarelli2017sustainable}. In this vein, building resilience into the critical infrastructure systems is becoming one of the prioritized concerns for planning sustainable cities. However, owing to multi-dimensional characteristics in transportation resilience, it remains as a difficulty to comprehensively measure this compound system property through abundant data. Although consensus on how to measure infrastructure resilience has never been achieved, we think the proposed BNM is a promising tool for providing a systematic analysis and systemic thinking to the problem, instead of finding a one-size-fits-all quantification criterion for evaluating transportation resilience. From the study, several rational implications can be discussed. 

As the four case studies demonstrated, the transportation systems' probabilities of being positively resilient are overall at a medium level, suggesting that these megacities might not be as resilient as we expected. It nevertheless indicates that there is much room for future improvement and we indeed observed increasing trends in all cases. From the results, we are not able to acquire precise explanations for the ``V" shapes found in Beijing and Tianjin cases. Such an interesting shape might be triggered by certain events or political strategies. But again, the causal relationships could be extremely difficult to be pin-pointed. A similar constraint can also be recognized in the results of resilience-economic coupling analysis. However, according to China's development strategy plan in national transportation, we are optimistic about building future resilient transportation infrastructure in these cities.

To achieve this, we recommend that relevant transportation practitioners, such as designers, constructor, operators, and managers, need to work closely together. In terms of building resilience into our transportation infrastructure, special attention should be placed on how to quickly rebuild the infrastructure and make it more adaptive to future threats. Such considerations should be incorporated into the entire life cycle of transportation systems, including the design phase, the construction phase, and the operation and maintenance phase. As a type of modern socio-technical systems, progressive innovation of urban transportation should also be taken into account at the same time. It is not a static consideration as each stage plays an important and positive role. In other words, building resilience is not just a top-level master design at the initial stage, but also a progressive process and it requires the social system that manages, operates and maintains the infrastructure resilience to co-evolve during this process. On the one hand, existing knowledge about design and construction would define how well we could build and how robust is the infrastructure to maintain functionality. On the other hand, the resilience think integrates both social and technical aspects to form a compound and systems-oriented management strategy for coping with long-term sustainable development issues in our urban transportation systems.

Cities have different resilience levels in their infrastructure systems. As illustrated in the model, it is a multi-dimensional dynamic system property, which has a strong association with time and other factors such as economic growth (even though such association might not be an unchanging phenomenon). Decision-makers therefore often find that it is difficult to achieve trade-offs in many sustainable development problems, especially the social-technical imbalance issues in transportation systems. To tackle this problem, systemic thinkings should play an indispensable role in managing our transportation infrastructure. 

From a technical perspective, compared with traditional assessment metrics and models such as performance-based metrics (PMs) and topological network-based models (TNMs), the proposed BNM demonstrates a promising advantage in dealing with multi-dimensional and multi-facet assessment issues. PMs have advantages in its generality and user-friendly implementation. Tools in this category typically require time-series performance monitoring, which heavily relies on the proper selection of a key performance indicator (KPI) that can represent the entire system. TNMs can provide detailed diagnostic in the systems' topological structures, which is also a wide-applied category but is constrained to acquiring full knowledge of the system architecture. In contrast, the BNM tools are effective for multi-dimensional evaluation and, more importantly, its abductive inference feature enables a backward analysis for users, which is extremely useful for decision-making and problem detection.

One of the advantages of using qualitative methods to determine the CPTs in the proposed BNM is to maximize expert knowledge and professional assessment, which can provide a holistic and experience-based understanding for fuzzy and uncertain concepts such as infrastructure resilience~\citep{zheng2018development}. However, this can also become a limitation, which the study may over-rely on subjective information that might introduce individual-related bias. Furthermore, because infrastructure resilience still lacks a commonly agreed benchmark for its measurement, it makes the mathematical validation extremely difficult for the proposed BNM. Another limitation of this study would be the lack of comparisons. In this study, we only studied cities in China. How the resilience levels of these cities are if comparing to cities from other countries remains unknown. Also, even within the context of China, cities with different administrative levels should be compared to further test the model because there could be hidden relations among transportation resilience, regional economic development, and geographic characteristics. However, as an exploratory study, this study sheds new lights on the concepts of long-term multi-dimensional resilience and specific hazard-related resilience in the course of transportation infrastructure development.

\section{Conclusions}
This paper proposes a multi-dimensional systems-based Bayesian network model (BNM) for evaluating resilience in urban transportation systems. The model is constructed based logic reasoning of well-defined resilience frameworks and four cities in China are used as case studies to illustrate the applicability strength of the proposed model. Unlike hazard-specific resilience, this paper discusses the concept of long-term multi-dimensional resilience in urban transportation and investigates the relation between its resilience development and urban economic development. Sensitivity analysis and backward reasoning provide a detailed diagnosis of the model and identify basic factors that may play vital roles in evaluating resilience in urban transportation systems. Three conclusions can be summarized in this study.

\begin{itemize}
\item The overall level of transportation resilience in the selected cities, Beijing, Tianjin, Shanghai, and Chongqing, are moderate with values between 50\% to 60\% from 1998 to 2017. Although they all have an increasing trend in recent five years, Beijing and Tianjin demonstrate a clear ``V" shape, which indicates a strong dynamic characteristic of transportation resilience in long-term performance.

\item Compared with each city's economic development, we found that the resilience we evaluated in these cities' transportation systems does not always positively correlated with their economic growth throughout the 20-years observation, which demonstrates a multi-dimensional nature of the concept and its non-linear relationship with regional economic development.

\item The results obtained from the sensitivity analysis and abductive reasoning suggest that the capabilities to quickly rebuild and make changes to cope with future disturbance should be paid more attention by decision-makers in different stages of the transportation infrastructure, as they could be the driving forces for achieving more resilient mobility systems.
\end{itemize}

The study enriches the resilience assessment toolbox and provides a basis for decision-making and planning. The approach we demonstrated here could also be transferred to other urban systems. However, we also acknowledged limitations in this study and suggested to try to tackle them in future work. Unlike other traditional models, the proposed BNM does not offer a one-size-fits-all solution. It is a new attempt to study transportation resilience from a long-term sustainability perspective and we invite more studies, with different and alternative perspectives, to further develop this interesting topic.

\cleardoublepage
\singlespace
\section*{\small Acknowledgement}
This research was jointly conducted at the Future Resilient Systems at the Singapore-ETH Centre (established collaboratively between ETH Zurich and Singapore's National Research Foundation FI-370074011) and the Centre for Smart Infrastructure and Construction (CSIC) at the University of Cambridge.

\section*{\small Author contributions}
All authors contributed to the conception and design of this study, and have read and approved the final manuscript. 

\section*{\small Data access}
The row datasets are all open-source archives and available to the public by following the corresponding citations and references in the paper.

\section*{\small Conflicts of interest}
The authors declare no conflicts of interest.


\end{document}